\documentclass[aps,prl,twocolumn,superscriptaddress,showpacs]{revtex4}

\usepackage{graphicx}

\begin{document}

\title{Control of heat transport in quantum spin systems}

\author{Yonghong  Yan }
\affiliation{Department of Physics and Centre for Computational
Science and Engineering, National University of Singapore, Singapore
117542, Republic of Singapore}
\author{Chang-Qin Wu}
\affiliation{Department of Physics, Fudan University, Shanghai
200433, People's Republic of China} \affiliation{Department of
Physics and Centre for Computational Science and Engineering,
National University of Singapore, Singapore 117542, Republic of
Singapore}

\author{Baowen  Li}
\email[Corresponding author. Email: ] {phylibw@nus.edu.sg}
\affiliation{Department of Physics and Centre for Computational
Science and Engineering, National University of Singapore, Singapore
117542, Republic of Singapore} \affiliation{NUS Graduate School for
Integrative Sciences and Engineering, Singapore 117597, Republic of
Singapore}

\date{21 July 2008}

\begin{abstract}
We study heat transport in quantum spin systems analytically and
numerically. First, we demonstrate that heat current through a
two-level quantum spin system can be modulated from zero to a finite
value by tuning a magnetic field. Second, we show that a spin
system, consisting of two dissimilar parts - one is gapped and the
other is gapless, exhibits current rectification and negative
differential thermal resistance. Possible experimental realizations
by using molecular junctions or magnetic materials are discussed.

\end{abstract}

\maketitle

When an electron moves in solids, it carries charge, spin, as well
as heat. The charge degree, the basis of the modern electronics, has
been fully studied over the past decades and it is still being under
intensive investigation at the molecular level
\cite{Datta05,Nitzan03}. On the other hand, the spin degree of
freedom, which may  be utilized to carry and process information as
well, has also been explored intensively in past years, which has
resulted in an emerging field ``spintronics'' \cite{Wolf,RMP04}. In
this context, spin (magnetization) current in insulating magnetic
materials has attracted considerable attention \cite{Meier,Sentef}.
However, heat current in spin systems has not been well studied as
it should be.

In fact, with the rapid miniaturization and increase of operational
speed of microelectronic devices, a great amount of redundant heat
are produced, which will in turn affect device performance. Thus
heat dissipation and heat management is becoming more and more
important \cite{Schulze,Galperin}. Moreover, it was found recently
that heat due to phonons can be used to carry and process
information \cite{WangLi08}. Therefore, the study of heat conduction
is not only of fundamental important but also helpful for the design
and fabrication of heat dissipator and phononic devices. Indeed,
many interesting conceptual devices such as thermal rectifier
(diode) \cite{Terraneo,Li04,Eckmann,Saito06}, thermal transistor
\cite{Li06,Ojanen}, and thermal logic gate \cite{Li07} have been
proposed. Experimentally, a nanoscale solid state thermal rectifier
using deposited carbon nanotubes has been realized recently
\cite{Chang}, and a heat transistor - control heat current of
electrons- controlled by a voltage gate has also been reported
\cite{Saira}. Furthermore, Segal and Nizan showed that thermal
rectification can appear in a two-level system asymmetrically
coupled to phonon baths \cite{Segal,Segal05b}, providing  the
possibility to control heat at a microscopic level.

In this Letter,  we demonstrate that the heat part from spins can be
also modulated and controlled. For example, we show that heat
current in a two-level system can be modulated from zero to a finite
value by tuning the magnetic field, $h$. Near $h=0$, the two levels
are almost degenerate and the system can jump easily between them;
thus, heat current (proportional to $h^2$) is small. As $h$
increases from zero to a finite value, heat current increases
accordingly. We further consider a spin-$1/2$ system consists of two
different segments: one is gapped and the other is gapless. We show
that in such a structure thermal rectifying efficiency can be very
high, up to 10 times. This system also exhibits negative
differential resistance, a feature which is necessary for building
up a thermal transistor.

\begin{figure}
\centering
\includegraphics[angle=0,scale=1.0]{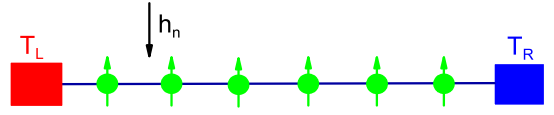}
\caption{(color online) A schematic representation of the model with
size $N=6$. The spin model is connected to two phonon  baths held at
different temperatures, $T_L$ and $T_R$. An inhomogeneous field is
applied to introduce  an asymmetric structure. \label{model}}
\end{figure}

\emph{Model} For simplicity,  we consider an inhomogeneous
mesoscopic spin-$1/2$ chain whose Hamiltonian reads
\begin{eqnarray}\label{HAMILT}
H&=&\sum_{n=1}^{N}h_n\sigma_n^z-Q\sum_{n=1}^{N-1}(\sigma_n^x
\sigma_{n+1}^x+ \sigma_n^z \sigma_{n+1}^z),
\end{eqnarray}
where $N$ is the number of spins,  the operators $\sigma_n^x$ and
$\sigma_n^z$ are the Pauli matrices for the $n$th spin,  $Q$ is the
coupling constant between the nearest neighbor spins, and  $h_n$ is
the magnetic field strength  (Zeeman splitting) at  the $n$th site.
Fig.~\ref{model} shows a schematic representation of this model. In
the thermodynamic limit, the model is an analog of the Heisenberg
spin-$1/2$ chain, wherein  the heat conduction is ballistic
\cite{Zotos97,Sologubenko00}.  At the microscopic level,
(\ref{HAMILT}) may  be viewed as a simplified version of a molecular
model in low temperatures  \cite{Segal05b}.

We use the quantum  master equation (QME) to study heat conduction
in this model. Two phononic  baths of different temperatures are
connected to the system at the ends. By tracing out the baths within
the Born-Markov approximation, we obtain the equation of motion for
the reduced density matrix of the system ($\hbar=1$)
\cite{Saito,Kubo},
\begin{eqnarray}\label{eq:qme}
   \frac{d\rho}{dt}=-i[H,\rho]+\mathcal{L}_L\rho+\mathcal{L}_R\rho,
\end{eqnarray}
where $\mathcal{L}_L\rho$  ($\mathcal{L}_R\rho$)  is  a dissipative
term due to the coupling with the left (right) bath.
$\mathcal{L}_L\rho$ is given by $
\mathcal{L}_L\rho=\left([X_L\rho,\sigma^{x}_1]+\mathrm{H.c.}\right)$,
and  $\mathcal{L}_R\rho$ can be given in a similar way. Here the
operator $X_L$ is defined through
\begin{eqnarray}
\langle
m|X_L|n\rangle=\lambda\varepsilon_{m,n}n_L(\varepsilon_{m,n})\langle
m|\sigma^{x}_1|n\rangle,
\end{eqnarray}
where $\varepsilon_{m,n}\equiv\varepsilon_{m}-\varepsilon_{n}$,
 and $n_L(\varepsilon_{m,n})=\left(e^{\varepsilon_{m,n}/T_L}-1\right)^{-1}$
 is the Bose distribution ($k_B=1$) with $T_L$ being the temperature of left bath.
  \{$|n\rangle$\} and \{$\varepsilon_{n}$\}
are the eigenstates and eigenvalues of the system Hamiltonian $H$,
respectively. $\lambda$ is the coupling strength with the bath. The
bath spectral function we used is an Ohmic-type. The master equation
(\ref{eq:qme}) is solved by the fourth-order Runge-Kutta method. In
numerical simulations, we take $\lambda=0.01$;  the simulation time
is  chosen long enough such that the final density matrix $\rho$
reaches a steady state.

\begin{figure}
\centering
\includegraphics[angle=0,scale=0.70]{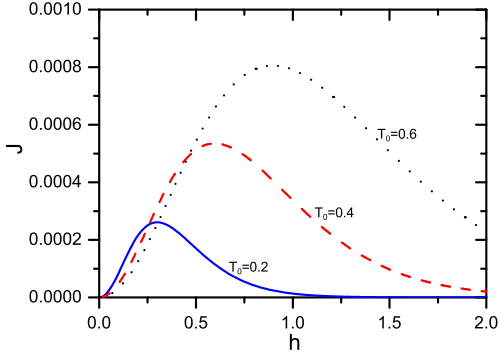}
\caption{(color online){\bf Modulation of Heat Current}. Heat
current as a function of the field. The bath temperatures are
$T_L=T_0+0.05$ and $T_R=T_0-0.05$. Three cases are shown: $T_0=0.2$
(solid line), $T_0=0.4$ (dashed line), and $T_0=0.6$ (dotted line).
\label{tuning}}
\end{figure}

 \emph{Modulation of heat current} To demonstrate the modulation of heat current by
 an external field, we consider a simple two-level system (TLS),
 namely, $N=1$; thus Eq.
 (\ref{HAMILT}) becomes
$H=h\sigma^z$.  From the QME, we obtain heat current,
\begin{eqnarray}\label{current:tls}
J=\lambda (2h)^2\frac{n_L(2h)-n_R(2h)}{1+n_L(2h)+n_R(2h)}.
\end{eqnarray}
Fig.~\ref{tuning} shows heat current as a function of the field. The
temperatures of the left and right baths are $T_L=T_0+0.05$ and
$T_R=T_0-0.05$, respectively. $T_0$ is the mean temperature. In
Fig.~\ref{tuning}, we see that heat current first increases with the
field and then decreases. In low fields ($h\ll T_0$), $n_{L,R}\sim
T_{L,R}/h$, then  $J\propto h^2$. In high fields ($h\gg T_0$),
$n_{L,R}\sim e^{-2h/T_{L,R}}$, then $J\propto h^3e^{-2h/T_{0}}$,
implying that heat current decays to zero when $h$ is large.
Therefore, in such a model we can modulate the current from zero to
a finite value by gradually switching on the external field. This
result is similar to those observed in one dimensional spin-$1/2$
systems recently \cite{Sologubenko08,Sologubenko07}, although in our
case we consider just one spin. In Fig.~\ref{tuning}, we can also
see that the current can be modulated in a much wider region in a
high temperature, such as $T_0=0.6$.

   \begin{figure}
\centering
\includegraphics[angle=0,scale=1.10]{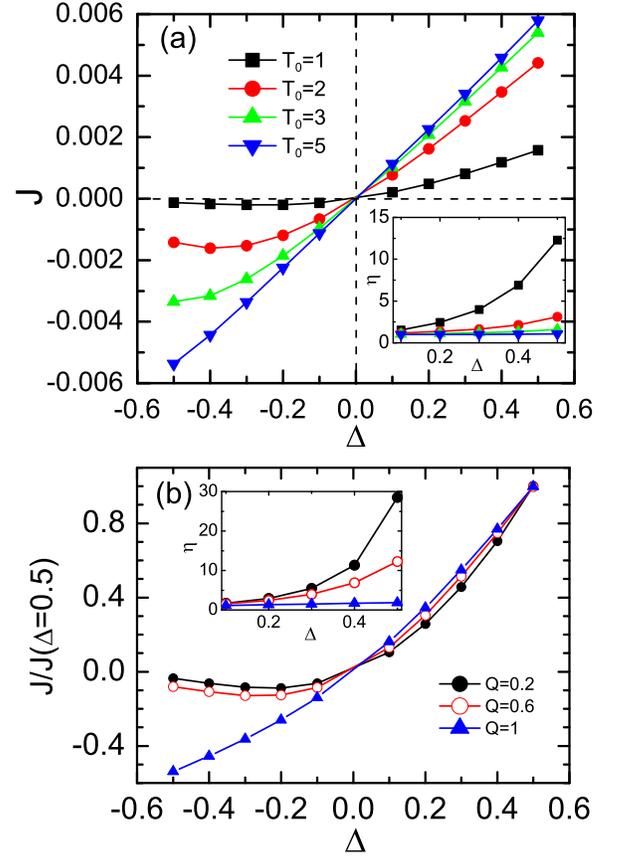}
\caption{(color online) {\bf Rectification of Heat Current}. (a)
Heat current vs temperature difference. The temperatures of the
baths are $T_L=T_0(1+\Delta)$ and $T_R=T_0(1-\Delta)$ where $T_0$ is
the mean temperature, ranging from   $T_0=1$ to $T_0=5$. $N=6$ and
$Q=0.6$. $h_n=1$ if $1\leq n\leq N/2$ and $h_n=0$ otherwise. (b)
Normalized heat current [$J/J(\Delta=0.5)$] vs $\Delta$ with
different couplings. Here $T_0=1$. The insets  show the
rectification efficiency, $\eta=|J_{+}/J_{-}|$, vs the temperature
difference. The lines are to guide the eye. \label{Q06a}}
\end{figure}

 \emph{Rectification of Heat Current} We now consider $N$ spins
 in an inhomogeneous magnetic
   field, namely,
 $h_n=1$ (as the unit of energy) if $1\leq n\leq N/2$ and $h_n=0$
otherwise (see Fig.~\ref{model}). The temperatures of the left and
right baths are $T_L=T_0(1+\Delta)$ and $T_R=T_0(1-\Delta)$,
respectively, where $T_0$ is a mean temperature and $\Delta$ is the
dimensionless temperature difference. The current operator may be
defined through the equation of continuity. In Fig.~\ref{Q06a}, we
show the heat current as a function of the temperature difference
with the mean temperature ranging from $T_0=1$
    to $T_0=5$.
   When the  temperature is low ($T_0=1$), we
   observe that for $\Delta>0$ the heat current increases
   with  $\Delta$, while in the region $\Delta<0$ the heat current
   remains   very small. Thus, our model
   exhibits  thermal rectification; namely, heat flows favorably in one direction.
    However, when  the
   temperature becomes high, such as $T_0=5$, the magnitude of heat current
   changes little as the  bath temperatures are exchanged. In this
   case the model cannot act as a good rectifier.
   Nevertheless, in a wide range of  temperature ($T_0\lesssim 3$), this model shows
   thermal rectifying  effect; the mechanism will be illustrated later.
   In Fig.~\ref{Q06a}(b), we show the heat current for a model with different couplings.
    We can see  that the rectifying  effect
   may sustain to large coupling constants. To quantify
   the rectification  efficiency,  we introduce the ratio,
   $\eta=|J_{+}/J_{-}|$,
   where $J_{+}$ is the current when $T_L>T_R$ and $J_{-}$ is the
   current when the temperatures are swapped, i.e., $T_L<T_R$.
   In a  weak coupling case, e.g., $Q=0.2$, the efficiency may be more than
   10 [see insets of Fig.~\ref{Q06a}(a) and (b)]; however, as the
   coupling $Q$ becomes stronger, the efficiency
   decreases.

\begin{figure}
\centering
\includegraphics[angle=0,scale=0.8]{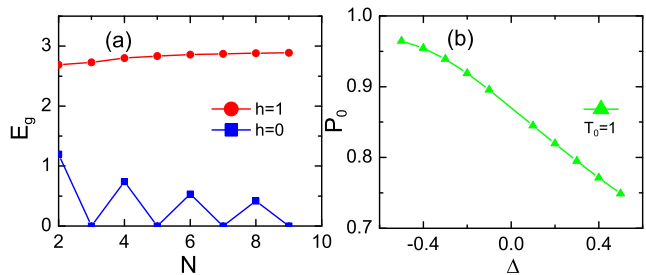}
\caption{(color online) (a) Energy gap of an isolated homogeneous
system ($h_n=h$) vs the  system size. The gap is  the energy
difference between the ground state  and the first excited state.
$Q=0.6$. (b) The probability, $P_0$,  to find the left part of the
system in the ground state
 as a function of $\Delta$.  $T_0=1$, $N=6$, $Q=0.6$, and
 $h_n=1$ if  $1\leq n\leq N/2$ and zero
otherwise. \label{Q06_occup}}
\end{figure}

To understand the rectifying effect qualitatively, we calculate the
energy gap $E_g$ for a homogeneous isolated system ($h_n=h$) and
show it in Fig.~\ref{Q06_occup}(a). The energy gap is obtained by
calculating the energy difference between the ground state  and the
first excited state. In a finite field $h=1$, we observe an energy
gap, $E_g\sim 2.8$, which is slightly dependent on the size. In the
absence of  the field ($h=0$), a small gap which vanishes in the
thermodynamic limit is seen. Note that if the size $N$ is odd, there
is no gap (degeneracy of the ground state); this is due to the
particle-hole symmetry if we map the spin model to a free Fermion
one by the Jordan-Wigner transformation.

At low temperatures, only a few low-lying states of each part are
relevant. Therefore, if the left part (with a gap)  is in contact
with a cold bath and  the right with a hot bath, i.e., $T_L=T_c$ and
$T_R=T_h$ ($T_c<T_h$), then the left part mainly remains in the
ground state. This is reflected in Fig.~\ref{Q06_occup}(b), which
shows the probability, $P_0$, to find the left part in the ground
state for different $\Delta$. In this case,  the transition rate of
the left part between different levels is low, and thus the heat
current is small although the right part (no gap or a small gap) may
jump easily between different levels (see also Fig.~\ref{tuning} for
reference). Reversely, if the left part of the system is in contact
with a  hot bath and the right with a cold one, the transition rate
of the left part between different levels is large, and then heat
current becomes large. This can also explain the low rectifying
efficiency when the mean temperature $T_0$ is increased. In this
case, the transition probability of the left part between different
levels may also be large; then, the magnitude of heat current
changes little when the bath temperatures are exchanged, implying a
low efficiency. In fact, we may observe thermal rectification
provided that $T_0\lesssim E_g$.

\begin{figure}
\centering
\includegraphics[angle=0,scale=0.8]{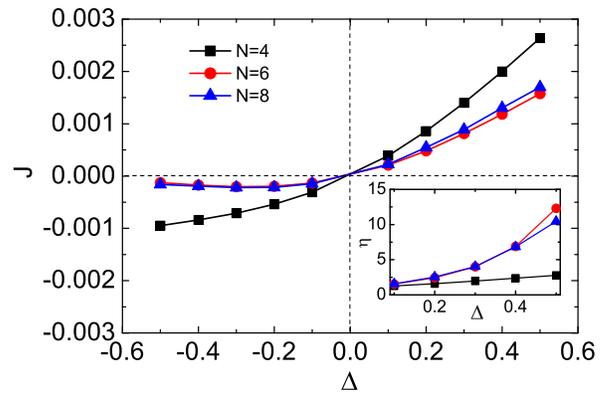}
\caption{(color online) {\bf Rectification of Heat Current}. Heat
current of a system with difference sizes. The bath temperatures are
$T_L=T_0(1+\Delta)$ and $T_R=T_0(1-\Delta)$. $T_0=1$ and $Q=0.6$.
The field is $h_n=1$ if $n\leq N/2$ and $h_n=0$ otherwise. The inset
shows the rectification efficiency  as a function of the temperature
difference.  \label{Q06_size} }
\end{figure}

In Fig.~\ref{Q06_size}, we show the heat current for a system with
different sizes. Here the coupling $Q=0.6$, and the field is $h_n=1$
if $n\leq N/2$ and $h_n=0$ otherwise. In the small size case
 ($N=4$), we just observe  thermal rectification with a low
efficiency. The reason may be  that  the effective coupling  between
the left and the right parts can be strong for a small size system.
As a result, the left part  may be excited by the right part even
though  it is connected to a cold bath. However, in a larger size
system, i.e., $N=6$ or $N=8$, we may observe both thermal
rectification  and negative differential resistance. Note also that
the efficiency changes very little when $N=6$ or $N=8$, implying
that  the model may  act as a rectifier  in an  even larger size
case.

\begin{figure}
\centering
\includegraphics[angle=0,scale=0.7]{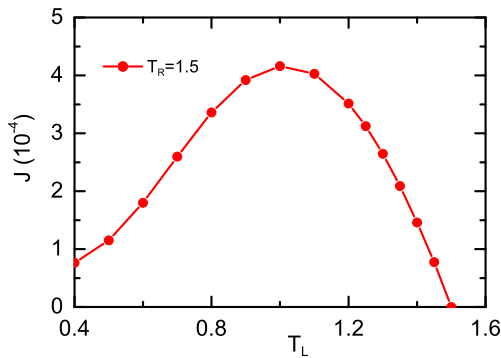}
\caption{(color online) {\bf Negative differential thermal
resistance (NDTR)}. Heat current vs temperature of left heat bath,
 $T_L$,  with fixed $T_R=1.5$. The other parameters are $Q=0.6$, $N=6$,
 $h_n=1$ if $1\leq n\leq N/2$ and $h_n=0$ otherwise. The NTDR is
 clearly seen for $T_L<1.0$.
\label{Q06_TRfix} }
\end{figure}

\emph{Negative Differential Thermal Resistance (NDTR)} In fact, in
Fig.~\ref{Q06a}, we can also observe NDTR in the region of
$\Delta<0$, i.e., the decrease of heat current with the increase of
temperature difference. A clearer representation is shown in
Fig.~\ref{Q06_TRfix}, where we fix  the temperature of the right
bath, $T_R=1.5$. We see that when the temperature of the left bath
$T_L$ is increased from $0.4$ to $1.0$, i.e., decreasing the
temperature difference, thermal current increases. The reason is
that: If $T_L$ is low, the left part is rarely excited, implying a
small current; otherwise, current may become large.  NDTR is an
important physical property that may be used to build spin-based
thermal transistors.

\emph{Summary} We have studied the possibilities to control heat
current in mesoscopic  spin models. We have showed that
 heat current could be modulated from zero to a finite value
  in a two-level system by tuning the
magnetic field. We have also studied  thermal rectification and
negative differential thermal resistance in an asymmetric model.
 The model consists of two parts: the left part is  gapped, and
the right part is gapless. Such a structure is of great importance
for the model to exhibit rectification and NDTR.   In certain cases,
we have found that the rectification efficiency, $|J_{+}/J_{-}|$,
can be larger than 10. Finally, we would like to discuss the
possible realizations of the model in experiment. The first is to
make use of the asymmetric structure in molecular bridges that can
be easily introduced. For example, we may use a molecule consisting
of two (weakly) coupled nonidentical spatially separated segments;
each is taken to be an anharmonic system, e.g., anharmonic
vibrations or molecular librations, where at low temperatures only
the lowest (two) quantum states are relevant (see Fig.~\ref{model}).
However, in this case, there are  gaps in both segments, so the
rectifying effect may be not so high.  The second possible way is to
use magnetic materials or molecular magnets \cite{Bogani}. The model
may be made up of two coupled magnetic materials: one is gapped and
the other is gapless. For the gapped material, one could use
spin-Peierls systems or introduce  a magnetic field  to open a gap
($h/Q>2$) \cite{Sologubenko07}.

\begin{acknowledgments}
Y.Y. thanks Lifa Zhang for valuable discussions.  The work was
supported in part by an ARF grant R-144-000-203-112 from the
Ministry of Education, Singapore, an endowment grant,
R-144-000-222-646 from NUS, and MOE of China (Grant No, B06011) and
NSFC.
\end{acknowledgments}

\end{document}